\documentclass[prd,aps,showpacs]{revtex4}

\newcommand{\bb}{\begin{eqnarray}}
\newcommand{\ee}{\end{eqnarray}}
\newcommand{\bsigma}{\mbox{\boldmath $\sigma$}}

\begin{document}

\title{ \bf Scattering of spin-polarized electron
in an Aharonov--Bohm potential}
\author{V. R. Khalilov$^1$ and Choon-Lin Ho$^2$}
\affiliation{    
$^1$Faculty of Physics, Moscow State University, 119899, Moscow,
Russia\\ $^2$Department of Physics, Tamkang University, Tamsui
25137, Taiwan, Republic of China}



\begin{abstract}
The scattering of spin-polarized electrons in an Aharonov--Bohm
vector potential is considered. We solve the Pauli equation in 3+1
dimensions taking into account explicitly the interaction between
the three-dimensional spin magnetic moment of electron and
magnetic field. Expressions for the scattering amplitude and the
cross section are obtained for spin-polarized electron scattered
off a flux tube of small radius. It is also shown that  bound
electron states cannot occur in this quantum system. The
scattering problem for the model of a flux tube of zero radius in
the Born approximation is briefly discussed.
\end{abstract}


\maketitle 

\section{Introduction}

The quantum Aharonov--Bohm (AB) effect, predicted by Aharonov and
Bohm \cite{1}, has been analyzed in various physical situations in
numerous works (see e.g., Ref. \cite{2}). It occurs when an
electron travels in a certain configuration of a vector potential
$A_{\mu}$ in which the corresponding magnetic flux is restricted
to a finite-radius ($R$) tube topologically equivalent to a
cylinder. When an electron travels in an AB potential the electron
wave function acquires a (topological) phase which could have
physical effects on the behavior of the electron, such as the
interference pattern in the two-slit experiment. The AB vector
potential can produce observable effects because the relative
(gauge invariant) phase of the electron wave function, correlated
with a nonvanishing  gauge vector potential in the domain where
the magnetic field vanishes, depends on the magnetic flux
\cite{khu}.

When the external field configuration has the cylindrical
symmetry, a natural assumption is that the relevant quantum
mechanical system is invariant along the symmetry ($z$) axis and
the system then becomes essentially two-dimensional in the $xy$
plane. So, the models applied to describe  AB effect can usually
be reduced to the (2+1)-dimensional ones.

The results of Ref. \cite{1} for nonrelativistic  case modified by
using the Dirac equation in 2+1 dimensions were applied to other
problems. Solutions of the two-component Dirac equation in the AB
potential were first discussed by Alford and Wilczek in Ref.
\cite{aw} in a study  of the interaction of cosmic strings with
matter.  Relativistic quantum AB effect was studied in Ref.
\cite{hkh} for the free and bound fermion states by means of exact
analytic solutions of the Dirac equation in 2+1 dimensions for a
combination of an AB potential and the Lorentz three-vector and
scalar Coulomb potentials.

Note that the usual four-component Dirac equation in 2+1
dimensions (in the absence of $z$ coordinate) decouples into two
uncoupled two-component Dirac equations for spin projection $s=+1$
and $s=-1$. Thus, the two-component Dirac  equation describes the
planar motion of relativistic electron having only one projection
of three-dimensional spin vector. The upper (``large'') and lower
(``small'') components of the two-component wave function are
interpreted in terms of positive- and negative-energy solutions of
the Dirac equation in 2+1 dimensions.

The scattering of spin-polarized fermions in an Aharonov-Bohm
potential was first considered in 2+1 dimensions by Hagen in
\cite{crh}. The particle spin in \cite{crh} is artificially
introduced into the two-component Dirac equation as a new
parameter (see, also, \cite{ggt}). The term including this new
parameter appears in the form of an additional delta-function
interaction of spin with magnetic field in the Dirac equation.
Solutions of the this Dirac equation were then interpreted for the
case of 3+1 dimensions. The magnetic field strength $H$ was taken
to have the form
\begin{equation}
H=-\frac{a}{R}\delta(r-R)\label{hagen}~, ~~a,~R:{\rm constants},
\end{equation}
and the vector potential in the Coulomb gauge is specified as
$$
A^0=0, \quad A_r=0,\quad A_{\varphi}=\frac{a}{r}, \quad r>R;
~~~{\bf A=0}, r<R.
$$
Even if the limit $R\to 0$ is taken at the end of the calculation,
it is seen that the above magnetic configuration does not quite
correspond to the real case because the magnetic field is equal to
zero inside the tube. So the question  still remain as to how the
interaction of the electron spin with the magnetic field of flux
tube can modify the known AB phenomenon.

In this paper we would like to consider the effect of spin in the
scattering process of a spin-polarized electron off an AB flux
tube with a small but finite radius.  In order to take account of
the interaction between the three-dimensional spin magnetic moment
of electron and magnetic field we shall use solutions of the Pauli
equation in 3+1 dimensions, which contains the corresponding spin
term explicitly.

This paper is organized as follows. In Sec. II we find the
scattering states of electrons in an AB potential, and give a
semiclassical argument for the scattering of spin-polarized
electrons in an AB potential. In Sec. III we determine the
scattering states of electrons for the realistic case when the
magnetic field is concentrated inside the cylindrical tube of a
small radius, and the scattering cross section for spin-polarized
electrons. Possibility of the existence of bound states in an AB
potential is briefly discussed in Sect. IV.  In Sec. V we briefly
consider the scattering problem for the model of a flux tube of
zero radius in the Born approximation. Sect. VI concludes the
paper. In the Appendix, we briefly discuss the scattering of a
spin-polarized Dirac electron in an AB potential in 2+1
dimensions.

\section{Scattering of a spin-polarized
electron off an AB flux tube of zero radius}

The problem of scattering of a spin-polarized electron described
by a Pauli equation off an AB flux tube of zero radius was studied
in \cite{khal}.  For completeness we give the main results here.

\subsection{Scattering states}

Consider an electron of mass $m$ and charge $e<0$ in an AB
potential, which is specified in Cartesian or cylindrical
coordinates as
 \bb
  A^0=0,\quad A_x=-\frac{By}{r^2},\quad
A_y=\frac{Bx}{r^2}, \quad A_z=0;
\nonumber\\
A^0=0,\quad A_r=0,\quad A_{\varphi}=\frac{B}{r}, \quad A_z=0,
\quad  B=\frac{\Phi}{2\pi}, \label{eight1}\\
\quad r=\sqrt{x^2+y^2}, \quad
\varphi=\arctan(y/x).\phantom{mmmmmmmmm} \nonumber
 \ee
This potential describes the magnetic field of an infinitely thin
solenoid with a finite magnetic flux $\Phi$ in the $z$ direction.
The magnetic field ${\bf H}$ is restricted to a flux tube of zero
radius \bb
 {\bf H}=(0,\,0,\,H)=\nabla\times {\bf A}= B\pi\delta({\bf r}).
\label{e1s}
\ee

To take into account of spin, we consider in this paper the Pauli
equation of a spinor $\Psi(t,{\bf r},z)$
 \bb
i\hbar\frac{\partial}{\partial t}\psi(t,{\bf r},z)={\cal
H}\psi(t,{\bf r},z),\quad {\bf r}=(x, y), \label{eq12} \ee where
the Hamiltonian ${\cal H}$ is \bb {\cal H} =
\frac{1}{2m}\left(-i\hbar\frac{\partial}{\partial
x}+\frac{eBy}{cr^2}\right)^2 +
\frac{1}{2m}\left(-i\hbar\frac{\partial}{\partial
y}-\frac{eBx}{cr^2}\right)^2
-\frac{\hbar^2}{2m}\frac{\partial^2}{\partial z^2} + M\sigma_3H.
\label{e12}
\ee
 Here $M=|e|\hbar/2mc$ is the Bohr magneton and
$\sigma_3$ is the Pauli spin matrix
$$
\sigma_3=\left(\begin{array}{cc}
1 & 0\\
0 &-1\\
\end{array}\right).
$$
The last term in (\ref{e12}) describes the interaction of the
electron spin with a magnetic field.

One seeks solutions of the Pauli equation in the form
 \bb
 \Psi(t,
r, \varphi, z) &=&\exp(-iEt/\hbar+ik_zz/\hbar) f(r, \varphi)\psi
\nonumber\\
&\equiv&
 \exp(-iEt/\hbar+ik_zz/\hbar)\sum\limits_{l=-\infty}^{\infty}F_l(r)\exp(il\varphi)
\psi, \label{three}
 \ee
where $E$ is the electron energy, $k_z$ is the wave number in the
$z$-direction, $l$ is an integer, and $\psi$ is a constant
two-spinor.  The function $F_l(r)$ satisfies
 \bb
 \left(\frac{\partial^2}{\partial
r^2}+\frac{1}{r}\frac{\partial}{\partial r} -k_z^2 - \frac{l^2}{r^2}
- \frac{2|e|Bl}{c\hbar r^2} - \frac{e^2B^2}{c^2\hbar^2 r^2} +
\frac{2Em}{\hbar^2}-\frac{|e|Bs}{c\hbar}\frac{\delta(r)}{r}\right)F_l(r)=0.
\label{e13}
 \ee
 Here the number $s=\pm 1$ characterizes the
electron spin projection on the $z$ axis. Eq. (\ref{e13}) is
satisfied by the function $F_l(r)$ everywhere except for the point
${\bf r}=0$. Hence, in the range $r>0$ the linearly-independent
solutions for $F_l(r)$ are expressed as
 \bb
 F_l(r)=
a_lJ_{\nu}(k_{\perp}r)+b_lJ_{-\nu}(k_{\perp}r),
 \label{regs12}
 \ee
where $J_{\nu}(k_{\perp}r)$ and $J_{-\nu}(k_{\perp}r)$ are the
usual Bessel functions, and
 \bb
 \nu =|l+\gamma|\ne 0,
\quad \gamma=|e|B/\hbar c>0, \quad k_{\perp} = \sqrt{2mE/\hbar^2 -
k_z^2}, \label{notion12}
 \ee
$a_l$ and $b_l$ are constants. For $\nu>0$ the regular solution is
 \bb
 F_l(r)=J_{\nu}(k_{\perp}r). \label{sol1s}
 \ee
 For $\nu=0$ the
linearly-independent solutions for $F_l(r)$ are the Bessel
($J_0(k_{\perp}r)$) and Neumann ($N_0(k_{\perp}r)$) functions.
Note that when $\nu$ is an integer the magnetic flux is quantized.

The electron wave function (\ref{three}) with $E>0$ belongs to the
continuous spectrum and describe the states of scattering.  For
$B=0$ one recovers the free electron solutions from
Eq.(\ref{sol1s}). One sees from Eqs. (\ref{e12}) and (\ref{e13})
that the additional singular  ``potential'' acts only at the point
${\bf r}=0$ where the regular solution $J_{\nu}(k_{\perp}r)$ goes
to zero, so we conclude that the regular solutions are well valid
in the range $r>0$ if only new possible (bound) states do not
occur.

One can define the scattering amplitude in the conventional
manner. Assuming that the electron wave incidents from the left
along the $x$ axis. Hence, $k_z=0$ and $k_\perp$ reduces to
$k\equiv \sqrt{2mE}/\hbar$. The incident wave function is
$\psi=e^{ikx}$. For this case $\varphi$ is the scattering angle
measured from the positive $x$-axis. As $r\to \infty$, the
electron wave function have the asymptotic form
 \bb
 \psi_p(r, \varphi) = e^{ikx+i(|e|B/\hbar c)\varphi}
+\frac{f(\varphi)}{\sqrt{r}}e^{ikr}. \label{solscat}
 \ee
Here $f(\varphi)$ is the scattering amplitude. The scattering
amplitude is proportional to $S_l-1\equiv e^{2i\delta_l}-1$, where
$\delta_l=(\nu-l)\pi= \pi\gamma$ are the partial phase shifts.
They depend upon only the total magnetic flux. Let us write
$\gamma= n+\mu$, where $n$ is an integer and $0\le\mu<1$.  The
scattering amplitude is then found to be given by the AB formula
 \bb
 f_{\rm AB}(\varphi)=
\sqrt{\frac{i}{2\pi k}}\frac{e^{-i\varphi(n-1/2)}
\sin(\pi\gamma)}{\sin(\varphi/2)}. \label{ampscat2}
 \ee
This formula has also been obtained by Alford and Wilczek in Ref.
\cite{aw} (see, also \cite{hkh}) by solving the Dirac equation in
2+1 dimensions in an AB potential. Formula (\ref{ampscat2}) is the
scattering amplitude of unpolarized electrons, and is apparently
independent of the electron spin. Thus the corresponding cross
section is the same as that given by the known AB formula
 \bb
\frac{d\sigma}{d\varphi} =|f_{\rm AB}(\varphi)|^2=
\frac{\sin^2(\pi \gamma)} {2\pi k\sin^2\varphi/2}~.\label{secscat}
 \ee

\subsection{Inclusion of spin}

Let us now discuss how the inclusion of electron spin polarized
along the direction of the a unit vector $\mathbf{n}$ may modify
the scattering amplitude. In the range $r>0$ the spinor $\psi$ can
be determined from the following equation
 \bb
(\bsigma\cdot{\bf n})\psi^s=s\psi^s, \label{spin1}
 \ee
where $\bsigma$ are the Pauli matrices, ${\bf n}= (\sin\vartheta,
0, \cos\vartheta$) is the unit vector characterized by the polar
($\vartheta$) and azimuthal ($\phi=0$) angles with respect to the
fixed axes $x, y, z$, and $s=\pm 1$ is the number which
characterizes the electron spin projection on the direction of the
unit vector ${\bf n}$. The solution of Eq. (\ref{spin1}) is given
by
 \bb \psi^s =
\frac{1}{\sqrt{2}} \left(
\begin{array}{c}
\sqrt{1+s({\bf n}\cdot{\bf n}_z)}\\
-s\sqrt{1-s({\bf n}\cdot{\bf n}_z)}
\end{array}\right),
\label{four1a}
 \ee
 where $({\bf n}\cdot {\bf n}_z)$ is a scalar
product of ${\bf n}$ with the unit vector ${\bf n}_z$ along the
$z$-axis.

It follows from Eq.~(\ref{e13}) that for the scattering of
spin-polarized particles only the following part of the potential
depending on the $l$ and $s$
 \bb
 \frac{2|e|Bl}{c\hbar
 r^2}+\frac{|e|Bs}{c\hbar}\frac{\delta(r)}{r}
\label{essen}
 \ee
 is essential.  One sees from Eq. (\ref{essen}) that there
arises no an asymmetry in the scattering for any spin projections
$s=0,\pm 1$. This means that the spin projection is conserved and
the dependence on the particle spin in the cross section arises
only because the propagation direction of electron is changed
after the scattering.

The change in the propagation direction of electron is due to the
momentum transfer occurring in the scattering process. Consider
the case when the spin of the scattered electron is oriented along
the direction of the unit vector ${\bf n}$. Then, the constant
spinor $\psi^s$  can be rewritten as
 \bb \psi^1
=\left(\begin{array}{c}
e^{-i\phi/2}\cos(\vartheta/2)\\
e^{i\phi/2}\sin(\vartheta/2)
\end{array}\right)
\label{sp1} \ee
and
 \bb
 \psi^{-1} =\left(\begin{array}{c}
-e^{-i\phi/2}\sin(\vartheta/2)\\
e^{i\phi/2}\cos(\vartheta/2)
\end{array}\right).
\label{sp2}
 \ee

Putting $\phi=0$ or $\phi=\pi/2$ in the initial spin function
$\psi_i^s$, we must put $\phi=\varphi$ or $\phi=\varphi+\pi/2$ in
the final spin function $\psi_f^s$. Furthermore, expanding the
initial spin-up and down functions over the spin function in the
final state as follows
 \bb
 \psi_i^1=A\psi_f^1+B\psi_f^{-1}, \quad
\psi_i^{-1}=C\psi_f^1+D\psi_f^{-1}, \label{spin2}
 \ee
 one can easily find that
$A=D^*=\cos(\varphi/2)+i\cos\vartheta\sin(\varphi/2)$,
$C=B^*=i\sin\vartheta\sin(\varphi/2)$ for both cases. Note that
$\cos\vartheta$ and $\sin\vartheta$ can be related to the scalar
product ${\bf n}\cdot {\bf n}_z$ and vector product ${\bf n}\times
{\bf n}_z$ of the initial unit vector ${\bf n}$  and the unit
vector ${\bf n}_z$, respectively.

Since the spin orientation cannot be changed, the cross section
(\ref{secscat}) for spin-polarized electrons with the spin along
the direction of the unit vector ${\bf n}$ must be multiplied by
the additional factors $|A|^2$ or $|D|^2$. Thus, for these cases
the cross section is described by following equation (see, also,
\cite{khal})
 \bb
\frac{d\sigma}{d\varphi}= \frac{\sin^2(\pi \gamma)} {2\pi k}
\left(\frac{1}{\sin^2\varphi/2}-({\bf n}\times {\bf
n}_z)^2\right). \label{x-sec}
 \ee
Eq.~(\ref{x-sec}), up to the replacement of $\varphi$ by
$\varphi+\pi$, coincides with the expression for the cross section
of polarized beams given in \cite{crh}. The difference is due to
the fact that the scattering angle here is measured from the
positive $x$-axis, whereas in \cite{crh} it is measured from the
negative $x$-axis. It is amazing that our result, using the
$(3+1)$-dimensional Pauli equation with the flux tube
concentrating at the origin, happens to coincide with that given
in \cite{crh}, in which the particle spin is artificially
introduced into the Dirac equation in 2+1 dimensions as a new
parameter and the magnetic field is concentrated only to the
surface of a cylinder of radius $R$, as mentioned in the
Introduction (see Eq.~(\ref{hagen})).

\section{Scattering of a spin-polarized
electron off an AB flux tube of small radius}

We shall now give a first principle derivation of
Eq.~(\ref{x-sec}), using the $(3+1)$-dimensional Pauli equation,
for the case in which the electron spin is perpendicular to the AB
flux tube of small but finite radius.  Eq.~(\ref{x-sec}) for
general spin orientation will then be obtained by interpolating
the formula for this case and that for the case of spin parallel
to the flux tube.

\subsection{Scattering states}

 In realistic situation the magnetic field
is restricted to a flux tube of small radius $R$, so the vector
potential ${\bf A}$ in the cylindrical coordinates $r, \varphi,
 z$ inside the flux tube can be specified as
 \bb
A^0=0,\quad A_r=0, \quad A_{\varphi}=\frac{1}{2}Hr, \quad
A_z=0. \label{one}
 \ee
 In the range  $r>R$ the vector potential
${\bf A}$ is the AB potential
 \bb
 A^0=0, \quad A_r=0,\quad A_{\varphi}=\frac{B}{r},
\quad A_z=0, \label{two}
 \ee
where $2B=HR^2$. The magnetic field ${\bf H}$ in the flux tube is
 \bb
 {\bf H}=(0,\,0,\,H)=\nabla\times {\bf A}
\label{e1}
 \ee
and is equal to zero outside the tube.

The standard approach is to obtain solutions of the Pauli equation
in the range $r<R$ and $r>R$, and then to impose the matching
condition at $r=R$. As far as the spin projection on the $z$ axis
is conserved we can substitute the eigenvalue $s=\pm 1$ for the
operator $\sigma_3$ in the Hamiltonian (\ref{e12}). After such a
substitution the spin dependence of the wave function becomes
unessential and $\Psi$ may be treated as an usual function of
position (see, e.g. \cite{ll}). Seeking solutions of the Pauli
equation in fields (\ref{one}) and (\ref{two}) in the form
(\ref{three}), we arrive at the differential equation for $F_l(r)$
in the range $r<R$ in the form
 \bb
\left[\frac{\hbar^2}{2m}\left(\frac{\partial^2}{\partial
r^2}+\frac{1}{r}\frac{\partial}{\partial r}
-\frac{l^2}{r^2}-k_z^2\right) + E -
\frac{(eHr)^2}{8mc^2}+\frac{\hbar|e|H(l+s)}{2mc}\right]F_l(r)=0.
\label{en20}
 \ee
In the range $r>R$ solution $F_l(r)$ is determined by
Eq.(\ref{regs12}).

As we are interested in the situation in which the electron is
moving in the $xy$-plane, we shall put $k_z=0$ in what follows.
The positive-energy solution for $F_l(r)$ regular in the range
$r<R$ is
 \bb
 F_l(r)=A_le^{-x/2}x^{|l|/2}\Phi(a, d; x),
\label{regs}
 \ee
 where $A_l$ is a constant,
$x=|eH|r^2/2c\hbar\equiv r^2\gamma/R^2$, $\Phi(a, d; x)$ is the
confluent hypergeometric function, and
 \bb
a=-\left(\frac{EmR^2}{2\gamma \hbar^2}-\frac{|l|+l+s+1}{2}\right),
\quad \omega=\frac{|eH|}{mc}, \quad d=|l|+1. \label{param}
 \ee
 We emphasize that $a$ depends on the spin parameter $s$.
This solution is normalized for any $l$. A second
linearly-independent solution $F_l(r)$ in the range $r<R$
(irregular at $r=0$) for integers $d$ is given by the confluent
hypergeometric function $\Psi(a, d; x)$ in the form (see, for
example, \cite{GR})
 \bb
 F_l(r)=
B_le^{-x/2}x^{|l|/2}\Psi(a, d; x), \label{irregs}
 \ee
 It should be
noted that this solution is normalized only when $l=0$. Solutions
(\ref{regs}) and (\ref{irregs}) differ from those obtained in
\cite{crh}. This is because the magnetic field is equal to zero in
the range $r<R$ in the model used in \cite{crh}.

In the range $r>R$ the linearly-independent solutions for $F_l(r)$
are given by
 \bb
 F_l(r)= a_lJ_{\nu}(kr)+b_lJ_{-\nu}(kr).
\label{regs1} \ee
 For $\nu=0$ the linearly-independent solutions for $F_l(r)$
are the Bessel ($J_0(kr)$) and Neumann ($N_0(kr)$) functions. The
functions $F_l(r)$ are just the Fourier coefficients in the
expansion of the spatial wave function.

In the range $r<R$, owing to the term proportional to the
$\sigma_3$ matrix, the constant spinor $\psi$ is determined from
the equation
 \bb (\bsigma\cdot{\bf n}_z)\psi^s=s\psi^s,
\label{spin}
 \ee
 where $s=\pm 1$. Hence, the number $s=\pm 1$
characterizes the electron spin projection on the $z$ axis. The
solution of Eq. (\ref{spin}) has the form
 \bb \psi^s = \left(
\begin{array}{c}
\psi^1\\
\psi^{-1}
\end{array}\right) = \frac{1}{2}
\left( \begin{array}{c}
1+s\\
1-s
\end{array}\right).
\label{four}
 \ee
Thus, for $r<R$ the spatial wave function depends only on the spin
projection in (or opposite) the magnetic field. The form
(\ref{regs}) implies that we can consider in the range $r<R$ the
spatial wave function depending explicitly on the number $s$,
which selects a particular value as the spin projection on the $z$
axis. The number $s$ will explicitly appear in the continuity
relations for the spatial wave functions at the point $r=R$. One
sees from Eq. (\ref{four}) that the magnetic field concentrated in
the range $r<R$ can have physical effect on the spin state of the
electron if the electron spin does not lie exactly along the $z$
axis in the range $r>R$.

\subsection{Scattering cross section for spin-polarized electrons}

Now we must match the spatial wave functions and their derivatives
at the point $r=R$ using the continuity relations. Results,
obviously, depend upon the orientation of electron spin. Indeed,
if in the range $r>R$ the electron spin is oriented in a direction
of the flux tube then the continuity relations for the spatial
wave functions will not depend upon the number $s$ because the
electron spin can only be oriented along the $z$ axis inside the
flux tube and its projection is conserved. Therefore, when the
unit vectors ${\bf n}$ and ${\bf n}_z$ are parallel, we can put
$s=0$ in the spatial wave functions.
 But if the electron spin is not oriented in a
direction of the unit vector ${\bf n}_z$ then the spatial wave
functions depend on the number $s=\pm 1$. Hence, the continuity
relations can be written as
 \bb
  F_l(R-\delta)&=& F_l(R+\delta),~~
\delta\to 0, \label{cont2}\\
\left(\frac{dF_l(r-\delta, s)}{dr}\right)_{r=R, \delta\to 0} &=&
\left(\frac{dF_l(r+\delta)}{dr}\right)_{r=R, \delta\to 0},
\label{sew2}
 \ee
 where $s=\pm 1$.

Using the formula
 \bb
 \frac{d}{dx}\Phi(a, d;
x)=\frac{a}{d}\Phi(a+1, d+1; x)
 \ee
the continuity relations can be written as
 \bb
A_le^{-(\gamma/2)}(\gamma)^{|l|/2}\Phi(a, d; \gamma)=
a_lJ_{\nu}(kR)+b_lJ_{-\nu}(kR), \label{contfun}
 \\
A_l\frac{\gamma}{R}e^{-(\gamma/2)}(\gamma)^{|l|/2}\left[\left(\frac{|l|}
{\gamma}-1\right)\Phi(a, d; \gamma)+\frac{2a}{d}\Phi(a+1, d+1;
\gamma)\right]
=\nonumber\\
a_l\frac{d}{dR}J_{\nu}(kR)+b_l\frac{d}{dR}J_{-\nu}(kR),
\label{match1}
 \ee
The case $\gamma=|e|HR^2/2\hbar c<1$ is of interest only. Indeed,
we can always obtain this magnitude for $\gamma$ by choosing the
appropriate magnitude of $H$ or $R$.
 For $\gamma < 1$ we can keep only the leading terms in the
expansion of the confluent hypergeometric function $\Phi(a, d;
x)$, and the continuity relations for small $\gamma$ can be
written as \bb
 &&C_l\left[1+\left(\frac{a}{d}-\frac12 \right)\gamma\right]=
N_l(kR)^{|l+\gamma|}\left[1-\frac{kR}{2(|l+\gamma|+1)}+\frac{(kR)^2}
{8(|l+\gamma|+1)(|l+\gamma|+2)}\right]
\nonumber\\
&&+D_l(kR)^{-|l+\gamma|}\left[1-\frac{kR}{2(-|l+\gamma|+1)}+\frac{(kR)^2}
{8(-|l+\gamma|+1)(-|l+\gamma|+2)}\right], \label{funcR}
 \ee

\bb
&&C_l\left[|l|+\left(\frac{a}{d}-\frac{1}{2}\right)\gamma(|l|+2)\right]=
N_l(kR)^{|l+\gamma|}\left[|l+\gamma|-\frac{kR}{2}+\frac{(kR)^2}
{4(|l+\gamma|+1)}\right]
\nonumber\\
&&+D_l(kR)^{-|l+\gamma|}\left[-|l+\gamma|-\frac{kR}{2}-\frac{(kR)^2}
{4(|l+\gamma|-1)}\right], \label{derivR}\ee where \bb
C_l=A_l(\gamma)^{|l|/2}, \quad
N_l=\frac{a_l}{2^{|l+\gamma|}\Gamma(|l+\gamma|+1)}, \quad
D_l=\frac{b_l}{2^{-|l+\gamma|}\Gamma(-|l+\gamma|+1)}
\label{sew}\ee and $\Gamma(z)$ is the gamma function of argument
$z$.

Note that at $R\to 0$ the situation in which the electron is most
probably found near the origin can only occur when the
spin-depending potential is attractive,  which requires $\gamma
s<0$ and $l=0$. Putting in Eqs. (\ref{funcR}) and (\ref{derivR})
$l=0$, we obtain
 \bb
  C_0=N_0(kR)^{|\gamma|}+D_0(kR)^{-|\gamma|}
\label{funcRf}\ee and \bb C_0\frac{\gamma
s}{|\gamma|}=N_0(kR)^{|\gamma|}+D_0(kR)^{-|\gamma|},
\label{derivRf}
 \ee
 from which one easily finds
  \bb
N_0(kR)^{|\gamma|}&=&\frac12\left(1+\frac{s
\gamma}{|\gamma|}\right)C_0,\label{solu1}\\
 D_0(kR)^{-|\gamma|} &=&
\frac12\left(1-\frac{s\gamma}{|\gamma|}\right)C_0.
\label{solu2}
 \ee
For $\gamma>0$, $s=-1$, we have $N_0=0$, which implies that the
solution $F_0(r)\to J_{-|\gamma|}(kr)$ in the range $r>R$.

It should be emphasized again that the $s$ term occurs in the
continuity relations only when the electron spin is not oriented
in the $z$ direction for $r>R$. It follows from Eqs.
(\ref{contfun}) and (\ref{funcR}) that at $R\to 0$ all the
coefficients $D_l$ vanish when the electron spin lies in the $z$
axis  and only the regular functions $J_{\nu}(kr)$ are present in
the expansion of the spatial electron wave function in the range
$r>R$. Therefore, in case the electron spin is directed along the
$z$ axis in the range $r>R$ the spatial electron wave function
$\psi(r, \varphi)$ is given by
 \bb
 \psi(r, \varphi)=\sum\limits_{l=-\infty}^{\infty} N_lJ_{\nu}(kr)e^{il\varphi},
\label{exsolsc}
 \ee
 where
 \bb
N_l=e^{-i(\pi/2)|l+\gamma|}. \label{ampscat1}
 \ee
From this it is easily checked that the scattering amplitude is
given by the AB formula (\ref{ampscat2}).

Now we turn to the situation where the electron spin is
perpendicular to the flux tube.   We have derived that for
$1>\gamma>0, s=-1$ the coefficient $N_0$ is equal to zero. This
implies that the $J_{|\gamma|}(kr)$ term with $l=0$ has to be
absent in the expansion of the spatial wave function and
Eq.(\ref{exsolsc}) must be written as
 \bb
 \psi(r, \varphi)=\sum\limits_{l=-\infty}^{\infty}
N_lJ_{\nu}(kr)e^{il\varphi} + J_{-|\gamma|}(kr)e^{i\pi\gamma/2}.
\label{newsc}
 \ee
Here the summation is carried out with the omission of the $l=0$
term, i.e. the $J_{|\gamma|}(kr)$ term.   Now the scattering
amplitude is given by
 \bb
f(\varphi)= \sqrt{\frac{i}{2\pi k}}\frac{e^{i\varphi/2}
\sin(\pi\gamma)}{\sin(\varphi/2)} + \sqrt{\frac{-i}{2\pi k}}
\sin(\pi\gamma). \label{ampsct}
 \ee
The second term on the right of this equation is absent when the
electron spin lies in the $z$ axis. The corresponding cross
section is
 \bb
\frac{d\sigma}{d\varphi} = \frac{\sin^2(\pi \gamma)} {2\pi k}
\left(\frac{1}{\sin^2\varphi/2}-1\right). \label{secsct}
 \ee

The cross section for the case when the spin of the scattered
electron lies in the unit vector ${\bf n}$ can be obtained by
interpolating Eqs. (\ref{secscat}) and (\ref{secsct}), taking into
account the conservation of spin projection on the $z$ axis. In
this case the second term, i.e., the ``1" term, on the right of
Eq. (\ref{secsct}) must be replaced by the $1-({\bf n}\cdot {\bf
n}_z)^2=({\bf n}\times {\bf n}_z)^2$ term and the cross section
takes the form (\ref{x-sec}).

\section{First Born approximation of electron scattering off a
flux tube of zero radius}

Let us briefly discuss  the scattering of spin-polarized electrons
in an AB potential in the first Born approximation. As was shown
above the cross section does not depend on the radius of a flux
tube so we consider the case when the magnetic field is restricted
to a flux tube of zero radius.

If the effective delta potential is positive, the correction to
the scattering amplitude which may arise from the inclusion of the
additional potential is angle-independent and is very small
compared with the usual AB term for small scattering angle
$\varphi$.  This term can be estimated if one considers the
scattering of a free particle in the delta potential in the $xy$
plane.

This part of scattering amplitude can be obtained by the formula
\cite{ll}
 \bb
f(\varphi)=\sqrt{\frac{k}{2i\pi}}\int\left(e^{2i\delta_{ph}(x)}-1)\right)
e^{-2ikx\sin(\varphi/2)}dx, \label{scatpot}
 \ee
where
 \bb
\delta_{ph}(x)= - \frac{\pi |e|Bs\delta(x)}{4\hbar kc}
\label{phasepot}
 \ee
is the phase shift. In the first Born approximation the scattering
amplitude (\ref{scatpot}) can easily be obtained in the form
 \bb f(\varphi)=-\frac{\pi|e|Bs}{\sqrt{2ik\pi}\hbar c}.
\label{ampfin}
 \ee
One sees that the corresponding correction to the cross sections
is the same for both the attractive and repulsive potential and
for $s=\pm 1$.  This formula coincides with the first term of the
expansion on $\gamma$ of second term on the right of Eq.
(\ref{ampsct}).

We see that the scattering amplitude of particles in the
two-dimensional delta potential does not depend on the scattering
angle at any energy of particles. This result generalizes to the
two-dimensional case the known exact result about the equality of
amplitude for toward- and backward-scattering in an
one-dimensional repulsive delta potential for any energy of
particles (see, for instance, \cite{sf}).

\section{Possibility of bound electron state in
an Aharonov--Bohm potential}

In \cite{khal} it was shown that for an AB flux tube of zero
radius a bound state of the electron may exist if the magnetic
flux is suitably quantized.   It was also shown that the
occurrence of such bound state can modify the scattering states,
but the total cross-section is unaffected.

Now we would like discuss whether a bound electron state may occur
in the quantum system in which the flux tube has a finite radius.
The wave function of such a bound state must belong to the
negative energy spectrum, and corresponds to a probability density
which is concentrated near the flux tube at the point ${\bf r}=0$
and vanishes at large $r$. In the range $r>R$ there is only one
solution of Eq. (\ref{e13}) with $E<0$, which is expressed through
the MacDonald function $K_0(z)$ of argument
$z=r\sqrt{2m|E|/\hbar^2}$. This solution must be matched with the
corresponding negative energy solution in the range $r<R$ at the
point $r=R$.

In the range $r<R$ the solution, which could be matched with the
MacDonald function $K_0(z)$, is the confluent hypergeometric
function $\Psi(a, 1; x)$ with $l=0$. It is well to note that the
electron wave function can be normalized also only for $l=0$. At
small $x$, the function  $\Psi(a, 1; x)$ becomes
 \bb
  \Psi(a, 1; x)
\cong - \frac{1}{\Gamma(a)}(\ln x - {\bf C}), \label{psif}
 \ee
 where
$\Gamma(a)$ is the gamma function of argument $a=1-E/\hbar\omega$
for $s=-1$ and $a=-E/\hbar\omega$ for $s=1$, and ${\bf C}=0.577
\ldots $ is the Euler constant. For negative $E$ when the energy
of bound state is very small $E\to -0$  it follows from Eq.
(\ref{psif}) that $\Psi(a, 1; x)$ for $s=1$ becomes $\Psi(a\to 1,
1; x)\to \ln x$ and  $\Psi(a\to+0, 1; x)$ for $s=-1$ becomes
$\Psi(a\to 0, 1; x)\to a\ln x$.  But  $\Psi(a, 1; x)$ behaves as
 \bb
\Psi(a, 1; x) \cong - \frac{1}{(\Gamma(a))^2}e^{x}x^{a-1}\ln x
\label{psif1}
 \ee
at large $x$ and so for the necessary values of the parameter
$a\to +0, 1$ the wave functions cannot be matched. Hence, we
conclude that the bound electron state cannot exist if the AB flux
has a finite radius.

\section{Summary}

In this paper we have considered the scattering of spin-polarized
electrons in an AB vector potential. By solving the Pauli equation
in 3+1 dimensions, taking into account explicitly the interaction
between the three-dimensional spin magnetic moment of electron and
magnetic field, we obtain expressions for the scattering amplitude
and the cross section for spin-polarized electron scattered off a
flux tube of small radius. We have also shown that bound electron
states cannot occur in this quantum system. We have also computed
the scattering amplitude of spin-polarized electron scattered off
a flux tube of zero radius by the Born approximation, and have
shown that it is consistent with the exact result.

Other than the AB phase, another interesting phase in quantum
mechanics is the Aharonov--Casher phase. This is the phase
acquired by an electron propagating in an electric field when the
spin-orbit coupling due to the interaction of the moving magnetic
moment with the electric field needs to be taken into account
\cite{ahc,hemck}.  We shall consider the Aharonov--Casher effect
in the scattering of spin-polarized neutral fermions with
anomalous magnetic moment elsewhere.

\vskip 1truecm



\appendix* 
\section{Scattering of
spin-polarized Dirac electrons in 2+1 dimensions}

In this appendix we shall demonstrate that the results of
\cite{aw,crh} for the scattering of relativistic spin-one-half
electrons in $2+1$ dimensions and the corresponding results
presented here for the scattering of nonrelativistic spin-1/2
electrons in $3+1$ dimensions coincide. For this we follow the
approach of \cite{crh} by introducing the particle spin in the
two-component Dirac equation as a new parameter. But unlike the
treatment in \cite{crh} which assumes the magnetic field
configuration in Eq.~(\ref{hagen}) , we take the potential to be
given by Eqs.~(\ref{one}) and (\ref{two}) in the absence of a
third partial ($A_z$) component.  We shall adopt the units where
$c=\hbar=1$.

The Dirac equation for a particle of mass $m$ and charge $e\equiv
-e_0<0 ~(e_0>0)$ in 2+1 dimensions in the absence of a third
partial coordinate in the potential $A_{\mu}$ is
 \bb
(\gamma^{\mu}
\hat P_{\mu} - m)\Psi = 0. \label{1two}
 \ee
Here the Dirac $\gamma^{\mu}$ matrices are conveniently defined in
terms of the Pauli spin matrices as (see, \cite{crh,ggt})
 \bb
\gamma^0= \sigma_3,\quad \gamma^1=is\sigma_1,\quad
\gamma^2=i\sigma_2, \label{1spin}
 \ee
 and $s$ is a new parameter
characterizing twice the spin value $s=\pm 1$ for spin ``up" and
``down", respectively, $\hat P_{\mu} = -i\partial_{\mu} -
eA_{\mu}$ is the generalized electron momentum operator. The
corresponding Dirac Hamiltonian is
 \bb
{\cal H}=\sigma_1P_2-s\sigma_2P_1+\sigma_3m+eA_0.\label{diham}
 \ee

We seek solutions of Eq. (\ref{1two}) in the form \cite{kh}
 \bb
 \Psi(t,{\bf x}) = \frac{1}{\sqrt{2\pi}}\exp(-iEt+il\varphi)
\psi_l(r, \varphi)~, \label{three1}
 \ee
where $E$ is the electron energy, $l$ is an integer, and
$\psi_l(r, \varphi)$ is a two-component function ({\it i.e.} a
$2$-spinor)
 \bb \psi_l(r,
\varphi) = \left( \begin{array}{c}
f_l(r)\\
g_l(r)e^{is\varphi}
\end{array}\right).
\label{2-spinor}
 \ee

Taking into account the easily checked relations
\bb sP_1\pm
iP_2=-ie^{\pm is\varphi}\left[s\frac{\partial}{\partial r}\pm
\left(\frac{i}{r}\frac{\partial}{\partial
\varphi}-\frac{e_0B}{r}\right)\right],\label{1rel}\ee for the
functions $f_l(r)$ and $g_l(r)$ we obtain the equations  \bb
s{df_l\over dr}-{l+e_0B\over r}f_l+(E+m)g_l = 0,
\nonumber \\
s{dg_l\over dr}+{s+l+e_0B\over r}g_l-(E-m)f_l = 0, \label{1eq7}
\ee in the range $r>R$ and \bb s{df_l\over dr}-\left({l\over
r}+{e_0Hr\over 2}\right)f_l+(E+m)g_l = 0,
\nonumber \\
s{dg_l\over dr}+\left({l+s\over r}+{e_0Hr\over
2}\right)g_l-(E-m)f_l = 0, \label{2eq7} \ee in the range $r>R$.

 Eliminating, for instance $g_l(r)$, we obtain the differential equation
for $f_l(r)$  \bb \left(\frac{\partial^2}{\partial
r^2}+\frac{1}{r}\frac{\partial}{\partial r}+E^2-m^2
-\frac{l^2}{r^2}- \frac{(e_0Hr)^2}{4}-e_0H(l+s)\right)f_l(r)=0,
\label{1en20} \ee in the range $r<R$ and \bb
\left(\frac{\partial^2}{\partial
r^2}+\frac{1}{r}\frac{\partial}{\partial r} + E^2-m^2 -
\frac{(l+e_0B)^2}{r^2}\right)f_l(r)=0 \label{1e13} \ee in the
range $r>R$. It is seen that Eqs. (\ref{1en20}) and (\ref{1e13})
for a tube of small radius up to the trivial replacement
$E^2-m^2=2mE$ coincide  respectively with the Eq.~(\ref{en20}) for
the range $r<R$ and Eq.~(\ref{e13}) for the range $r>R$.

In the range $r>R$, the electron wave function   has the form \bb
 \Psi_p(r, \varphi)=e^{-iEt+il\varphi}\sqrt{\frac{\pi p}{2E}}
\left( \begin{array}{c}
\sqrt{E+m}J_{\nu}(pr)\\
\pm\sqrt{E-m}e^{is\varphi}J_{\nu\pm s}(pr)
\end{array}\right).
\label{sol1} \ee Here $p = \sqrt{E^2 - m^2}$, and $J_{\nu}(pr)$ is
the Bessel function. In the range $r>R$ the linearly-independent
solutions for $f_l(r)$ are
 \bb
  f_l(r)= a_lJ_{\nu}(pr)+b_lJ_{-\nu}(pr),
\label{regs12a} \ee where $J_{\nu}(pr)$ and $J_{-\nu}(pr))$ are
the usual Bessel functions and
 \bb
 \nu =|l+\gamma|\ne 0,
\quad \gamma=e_0B>0,
\label{notion12a}
\ee
$a_l$ and $b_l$ are
constants. For $\nu>0$ the regular solution is
 \bb
f_l(r)=J_{\nu}(pr). \label{sol1sa}
 \ee
For $\nu=0$ the linearly-independent solutions for $f_l(r)$ are
the Bessel ($J_0(pr)$) and Neumann ($N_0(pr)$) functions.

Assuming that the incident electron wave is moving from the left
to the right along the $x$-axis.  The upper component of the
incident wave is $\psi=e^{ipx}$.  The electron wave function in
the potential (\ref{eight1}) must have the asymptotic form \bb
\psi_p(r, \varphi) = \left(
\begin{array}{c}
1\\
-ip/(E+ m)
\end{array}\right)e^{ipx+ie_0B\varphi}
+ \left( \begin{array}{c}
1\\
ip/(E+ m)
\end{array}\right)\frac{f(\varphi)}{\sqrt{r}}e^{ipr}
\label{solscat1}
\ee
as $r\to \infty$. Here $f(\varphi)$ is the
scattering amplitude.

Writing $\psi(r, \varphi)$ in the form
\bb
 \psi(r, \varphi)=\sum\limits_{l=-\infty}^{\infty} A_lJ_{\nu}(pr)e^{il\varphi},
\label{exsolsc1}
 \ee
it is easy to show  that
  \bb
  A_l=e^{-i(\pi/2)|l+e_0B|}.
\label{ampscat1a}
 \ee
 The scattering amplitude is proportional to
$S_l-1\equiv e^{2i\delta_l}-1$, where
$\delta_l=(\nu-l)\pi=e_0B\pi\equiv e_0\Phi/2\hbar c$ are the
partial phase shifts. They depend upon only the total magnetic
flux $\Phi$.

The coefficient before the term $e^{ipr}/\sqrt{r}$ is the standard
AB amplitude for the scattering of nonrelativistic particles
 \bb
 f_{\rm AB}(\varphi)=
\frac{1}{\sqrt{2\pi pi}}\frac{e^{-i\varphi(n-1/2)}
\sin(e_0\Phi/2)}{\sin(\varphi/2)}. \label{ampscat2a}
 \ee
Here $e_0\Phi=2\pi n+2\pi\Delta$ where $n$ is an integer, and
$\quad -1/2\le\Delta\le 1/2$. Amplitude (\ref{ampscat2a}) was
first calculated in Ref. \cite{aw}. This part of the scattering
amplitude is unaffected by the spin parameter $s$.

The effect of spin, leading to Eq.~(\ref{x-sec}), can be
considered in full analogy with Sect. III of this paper for the
upper components.

\vskip 1truecm

\begin{acknowledgments}  

\vskip 0.5cm

 This paper was supported  by a Joint Research Project
of the National Science Council (NSC) of the Republic of China
under Grant No. NSC 95-2911-M-032-001-MY2 (C.L.H), and the Russian
Foundation for Basic Research (RFBR) under Grant No.
NSC-a-89500.2006.2 and NSC-RFBR No. 95WFD0400022 (contract No.
RP06N04) (V.R.K), and in part by the Program for Leading Russian
Scientific Schools (Grant No. NSh-5332.2006.2)(V.R.K.).

\end{acknowledgments}  

\end{document}